\documentclass{article} \usepackage{slashed,epsfig,amsmath,amssymb,enumitem} \usepackage[papersize={8.5in,11in}]{geometry}
   \usepackage{bm}
\usepackage{graphicx}
\newcommand{\be}{\begin{equation}}
\newcommand{\ee}{\end{equation}}
\title{Complex Riemannian spacetime and singularity-free black holes and cosmology}
\author{J. W. Moffat\\
Perimeter Institute for Theoretical Physics, Waterloo, Ontario N2L 2Y5, Canada\\
and\\
Department of Physics and Astronomy, University of Waterloo, Waterloo,\\
Ontario N2L 3G1, Canada}
\begin{document}
\maketitle


\begin{abstract}
An approach is presented to address singularities in general relativity using a complex Riemannian spacetime extension. We demonstrate how this method can be applied to both black hole and cosmological singularities, specifically focusing on the Schwarzschild and Kerr black holes and the Friedmann-Lema\^itre-Robertson-Walker (FLRW) Big Bang cosmology. By extending the relevant coordinates into the complex plane and carefully choosing integration contours, we show that it is possible to regularize these singularities, resulting in physically meaningful, singularity-free solutions when projected back onto real spacetime. The removal of the singularity at the Big Bang allows for a bounce cosmology. The approach offers a potential bridge between classical general relativity and quantum gravity effects, suggesting a way to resolve longstanding issues in gravitational physics without requiring a full theory of quantum gravity.
\end{abstract}

\section{Introduction}

General Relativity (GR) has been remarkably successful in describing gravitational phenomena across a wide range of scales, from the solar system to cosmological distances. However, the theory predicts its own downfall in the form of singularities - points where spacetime curvature becomes infinite and the laws of physics, as we understand them, break down. These singularities appear in two particularly important scenarios: at the center of black holes and at the beginning of the universe in Big Bang cosmology.

The presence of these singularities has long been viewed as a significant problem in theoretical physics, indicating the limits of classical GR and the need for a more fundamental theory, often presumed to be a theory of quantum gravity. However, a full theory of quantum gravity remains elusive, leaving us with the challenge of addressing these singularities within the framework of classical or semi-classical approaches.

In this paper, we present an approach to singularity resolution using complex Riemannian spacetime extensions~\cite{Moffat1,Moffat2,Moffat3, Kerr1,Guendelman,Afshordi}. This method, inspired by techniques in complex analysis and building upon previous work in analytic continuation of spacetime metrics, offers a way to regularize singularities without requiring a full quantum theory of gravity.

Our approach involves extending the relevant spacetime coordinates into the complex plane and carefully choosing integration contours that avoid singular points. By doing so, we can define new, regularized coordinates that, when projected back onto real spacetime, yield singularity-free solutions. This technique preserves the essential features of the original solutions while removing problematic infinities.

We apply this method to two fundamental scenarios in gravitational physics:

a) The Schwarzschild and Kerr black holes, addressing the central singularity at r = 0.
b) The Friedmann-Lema\^itre-Robertson-Walker (FLRW) cosmological model, tackling the initial singularity at $t = 0$.

In both cases, we demonstrate how the complex spacetime approach leads to regular solutions that can be interpreted as introducing a minimal length scale, potentially related to quantum gravity effects. 

The paper is structured as follows: Section 2 provides a detailed description of the complex Riemannian method and its mathematical foundations. Section 3 applies the method to the Schwarzschild black hole, while Section 4 applies the method to the Kerr black hole. Section 5 addresses the cosmological singularity in the FLRW model. Finally, Section 6 concludes with a summary of our findings and suggestions for future research directions.

\section{Complex Riemannian geometry}

A complex Riemannian spacetime manifold is an extension of the real spacetime manifold into the complex domain. In this framework, we consider a 4-dimensional complex manifold that is locally homeomorphic to an eight-dimensional spacetime. This manifold is equipped with a complex metric tensor $g_{\mu\nu}$ that is symmetric under the exchange of its indices. The coordinates on this manifold are complex-valued functions $z^\mu$, where 
$\mu=0,1,2,3$. The real spacetime is embedded in this complex manifold as a real slice or submanifold. The complex nature of this manifold allows for analytic continuation of solutions to Einstein's field equations, providing a richer structure that can be used to address singularities present in the real spacetime.

In complex Riemannian geometry, we extend the concepts of traditional Riemannian geometry to the complex domain while maintaining symmetry properties. The metric tensor $g_{\mu\nu}$ is a complex-valued, symmetric tensor. It can be expressed as:
\be
g_{\mu\nu}=s_{\mu\nu}+ia_{\mu\nu},
\ee
where $s_{\mu\nu}$ and $a_{\mu\nu}$ are real symmetric tensors. We choose units with the velocity of light $c=1$.

The line element is given by:
\be
du^2=g_{\mu\nu}dz^\mu dz^\nu.
\ee
This complex line element generalizes the concept of distance in the complex manifold. The affine connection $\Gamma^\lambda_{\mu\nu}$ is complex-valued and symmetric in its lower indices:
\be
\Gamma^\lambda_{\mu\nu}=S^\lambda_{\mu\nu}+iA^\lambda_{\mu\nu},
\ee
where $S^\lambda_{\mu\nu}$ and $A^\lambda_{\mu\nu}$ are real and symmetric in $\mu$ and $\nu$.

The Riemann curvature tensor is defined analogously to its real counterpart but with complex-valued components:
\be
R^\rho_{\sigma\mu\nu}= B^\rho_{\sigma\mu\nu}
+ iI^\rho_{\sigma\mu\nu}=\partial_\mu\Gamma^\rho_{\nu\sigma} - \partial_\nu\Gamma^\rho_{\mu\nu} +\Gamma^\rho_{\mu\lambda}
\Gamma^\lambda_{\nu\sigma} - \Gamma^\rho_{\nu\lambda}
\Gamma^\lambda_{\mu\sigma}.
\ee

The complex symmetric approach allows for a richer structure and for a powerful tool for exploring the structure of spacetime beyond the limitations of real spacetime. By carefully choosing contours in the complex manifold, we can avoid singularities that appear in the real spacetime while maintaining the essential physical properties of the solutions. It offers insights into the nature of spacetime at extreme conditions where quantum gravity effects are expected to become important.

In the following sections, we will apply this complex Riemannian geometry to specific problems in general relativity, demonstrating its utility in addressing longstanding issues in gravitational physics. The projection of the complex symmetric geodesic equation onto physical real spacetime is a crucial step in connecting the mathematical framework of complex spacetime to observable physics. This process involves several steps and considerations. 

The geodesic equation in complex symmetric spacetime is given by
\be
\frac{d^2z^\lambda}{du^2}+\Gamma^\lambda_{\mu\nu}\frac{dz^\mu}{du}\frac{dz^\nu}{du}=0.
\ee
Here, $z^\lambda$ are complex coordinates and
u is a complex affine parameter. 

We decompose each complex quantity into its real and imaginary parts $z^\lambda=x^\lambda+iy^\lambda$ and u=s+iw.
Substituting these decompositions into the geodesic equation and separating real and imaginary parts, we get two coupled equations. The real part is given by
\be
\frac{d^2x^\lambda}{ds^2}
- \frac{d^2y^\lambda}{dw}+S^\lambda_{\mu\nu}\left(\frac{dx^\mu}{ds}\frac{dx^\nu}{ds}-\frac{dy^\mu}{dw}\frac{dy^\nu}{dw}\right)-
A^\lambda_{\mu\nu}\left(\frac{dx^\mu}{ds}\frac{dy^\nu}{dw}+\frac{dy^\mu}{dw}\frac{dx^\nu}{ds}\right)=0,
\ee
and the imaginary part is
\be
\frac{d^2y^\lambda}{ds^2}+\frac{d^2x^\lambda}{dw^2}+S^\lambda_{\mu\nu}
\left(\frac{dx^\mu}{ds}\frac{dy^\nu}{dw}+\frac{dy^\mu}{dw}\frac{dx^\nu}{ds}\right)+A^\lambda_{\mu\nu}
\left(\frac{dx^\mu}{ds}\frac{dx^\nu}{ds}-\frac{dy^\mu}{dw}\frac{dy^\nu}{dw}\right)=0.
\ee
    
To project onto real spacetime, we consider the limit where the imaginary parts of the coordinates and the affine parameter approach zero $y^\lambda\rightarrow 0$ and 
$w\rightarrow 0$. In this limit, the imaginary part equation becomes trivial, and the real part equation simplifies to:
\be
\frac{d^2x^\lambda}{ds^2}+S^\lambda_{\mu\nu}\frac{dx^\mu}{ds}\frac{dx^\nu}{ds}=0.
\ee
This is the standard geodesic equation in real spacetime, with $S^\lambda_{\mu\nu}$ being the real part of the complex connection. The real part of the complex connection, $S^\lambda_{\mu\nu}$, is not necessarily the same as the connection in the original real spacetime. It contains information from the complex extension and regularization process. This modified connection is what gives rise to the regularized, singularity-free behavior in the projected real spacetime.

The resulting geodesic equation in real spacetime describes the motion of particles in the regularized geometry. Trajectories remain smooth even in regions where singularities existed in the original spacetime. The curvature experienced by particles remains finite everywhere. Important conservation laws, such as energy, angular momentum are preserved in the projection process. The differences between geodesics in the original and regularized spacetime become significant only near the former singular regions. Outside the black hole horizons, the behavior closely approximates that of the original real physical spacetime, ensuring consistency with existing observations.

The projection of complex symmetric geodesics onto real spacetime provides a bridge between the mathematical framework of complex extension and observable physics. It results in regularized trajectories that avoid singularities, while preserving the essential physical features of the original spacetime. This process offers insights into how quantum gravity effects might manifest in classical geometries, providing a valuable tool for exploring extreme gravitational phenomena.

\section{Singularity-free Schwarzschild black hole in complex spacetime and physical spacetime}

To demonstrate how the complex, regular singularity-free black hole Schwarzschild solution can be projected onto the real spacetime to give a physical regular singularity-free black hole, we need to consider a contour integral in the complex plane that avoids the singularity. This approach allows us to regularize the singularity in the real spacetime.

The complex Schwarzschild metric is given by
\begin{align}
du^2=-f(\zeta)d\tau^2+\frac{d\zeta^2}{f(\zeta)}+\zeta^2(d\theta^2+\sin^2\theta d\phi^2),
\end{align}
where $u=s+iw$ is the complex proper time, $\tau$ is the complex coordinate time, $f(\zeta) = 1-2GM/\zeta$ and  $\zeta=r+i\kappa$, $\theta$, $\phi$ are complex spherical polar coordinates. The complex Schwarzschild metric is a solution of the complex vacuum field equations:
\be
\label{vacuumequations}
R_{\mu\nu}=0.
\ee

We define a new radial coordinate R($\zeta$) through a contour integration:
\be
R(\zeta)=\oint_C\frac{d\zeta}{\sqrt{f(\zeta)}}.
\ee
The contour C is chosen to avoid the singularity at $\zeta=0$. Evaluating this integral yields:
\be
\label{Rsolution}
R(\zeta)=\int_C\frac{d\zeta}{\sqrt{1-\frac{2GM}{\zeta}}}=\zeta\sqrt{1-\frac{2GM}{\zeta}}
+2GM\ln\left(\frac{\sqrt{\zeta}+\sqrt{\zeta-2GM}}{\sqrt{2GM}}\right).
\ee 
We can now express the metric in terms $R(\zeta)$:
\be
\label{metric}
du^2=-\left(1-\frac{2GM}{R(\zeta)}\right)d\tau^2+\frac{dR(\zeta)^2}{\left(1-\frac{2GM}{R(\zeta)}\right)}
+R(\zeta)^2(d\theta^2+\sin^2\theta d\phi^2).
\ee
The contour C is specifically chosen to avoid the singularity at $\zeta = 0$. The function $R(\zeta)$ is non-zero for all finite values of $\zeta$ and R(0) is excluded by the contour integration.

\begin{figure}
    \centering
    \includegraphics[width=0.5\linewidth]{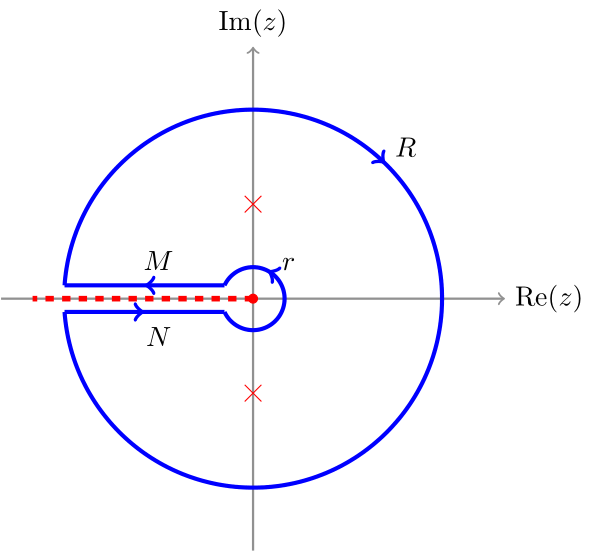}
    \caption{Complex plane contour integration for $f(\zeta)$ removing singularity at $\zeta=z=0$. Wikipedia.}
    \label{fig:enter-label}
\end{figure}

To create a complete maximally analytic manifold, we can introduce Kruskal-Szekeres type coordinates~\cite{Kruskal,Szekeres}. We define a tortoise coordinate:
\be
r_* = \int \frac{dr}{1-\frac{2GM}{R(\zeta)}},
\ee
and introduce null coordinates:
\be
u = t - r_*, \quad v = t + r_*.
\ee
Now we define Kruskal-Szekeres like coordinates:
\be
U = -\exp (-\kappa u), \quad V 
= \exp (\kappa v),
\ee
where $\kappa$ is the surface gravity. The metric in these coordinates takes the form:
\be
du^2 = -F(r)dUdV + R(\zeta)^2(d\theta^2 + \sin^2\theta d\phi^2).
\ee

To find F(U,V), we need to express it in terms of $R(\zeta)$. The general form will be:
\be
\label{Kruskalregularized}
F(U,V) = \frac{4}{\kappa^2}\left(1-\frac{2GM}{R(\zeta)}\right)
\exp\left(-\frac{R(\zeta)}{2GM}\right).
\ee
The exact form of $F(U,V)$ from our regularization procedure will be determined by $R(\zeta)$.

These Kruskal-Szekeres type coordinates for the regularized black hole will cover the entire black hole spacetime in a single coordinate patch and make the metric regular at both $\zeta = 0$ and $\zeta = 2GM$. They will also allow for a smooth extension through the event horizon, and reveal the maximal analytic extension of the spacetime, including regions that were hidden in the original coordinates. In this coordinate system, the event horizon is represented by the lines $U = 0$ or $V = 0$, and the spacetime is manifestly regular there. This completes the regularization process, providing a description of the black hole that is free of coordinate singularities and reveals the full analytic structure of the spacetime.

The Kretschmann scalar invariant is defined by
\be
K=R_{\alpha\beta\gamma\delta}R^{\alpha\beta\gamma\delta},
\ee
where $R_{\alpha\beta\gamma\delta}$ is the Riemann curvature tensor. The Kretschmann scalar $K$ for the regularized Schwarzschild metric is given by
\be
K=\frac{48G^2M^2}{R(\zeta)^6},
\ee
This result shows that the Kretschmann scalar remains finite for all values of $R(\zeta)$. This demonstrates that the complex extension and contour integration approach has successfully removed the singularity in the Schwarzschild black hole, while preserving the essential features of the black hole solution outside the black hole and at large distances.

\section{Singularity-free Kerr black hole in complex spacetime and physical spacetime}

The complex spacetime approach for removing the singularity in the Schwarzschild black hole can be extended to the Kerr black hole solution, although the process is more intricate due to the additional complexity of the Kerr metric. First, recall the Kerr metric in Boyer-Lindquist coordinates~\cite{Kerr}:
\begin{align}
du^2=-\left(1-\frac{2GMr}{\Sigma}\right)dt^2-\frac{4GMar\sin^2\theta}{\Sigma} dtd\phi+\frac{\Sigma}{\Delta}dr^2+\Sigma d\theta^2+\left(r^2+a^2+\frac{2GMa^2r\sin^2\theta}{\Sigma}\right)\sin^2\theta d\phi^2,
\end{align}
where $\Sigma=r^2+a^2\cos^2\theta$ and 
$\Delta=r^2-2GMr+a^2$. The Kerr solution has two singularities. A coordinate singularity at $\Delta=0$, which gives the inner and outer event horizons. A ring singularity at $\Sigma=0$, which occurs when $r=0$ and $\theta=\pi/2$.

To apply the complex symmetric approach, we extend the radial coordinate into the complex plane. We can define a contour C that depends on both $\zeta$ and $\theta$:
\be
\zeta(\lambda,\theta)=r(\lambda,\theta)+il (\lambda,\theta),
\ee
where $\lambda$ is a real parameter and 
$l (\lambda,\theta)$ is chosen to avoid the singularities for all $\theta$. We then define a new radial coordinate R along this contour:
\be
R(\zeta)=\oint_C\frac{d\zeta}{\sqrt{\Sigma}}.
\ee

The explicit evaluation of this integral is more complex than in the Schwarzschild case and may require numerical methods for general values of the parameters. The metric can then be rewritten in terms of $R(\zeta)$, $\theta$, $\tau$, and $\phi$. The resulting metric will be regular at the original singularities. The physical interpretation is similar to the Schwarzschild case. The ring singularity is replaced by a regular region of finite size, which can be interpreted as a quantum gravity effect.

The horizon structure of the Kerr black hole (inner and outer horizons) is preserved in this regularized version, but the central singularity is removed. An important consideration in the Kerr case is the preservation of the symmetries of the original metric, particularly the axial symmetry represented by the Killing vector 
$\partial\phi$. The complex extension must also preserve the asymptotic behavior of the Kerr metric, ensuring that it reduces to the Minkowski metric at large distances.
This extension to the Kerr metric demonstrates the power and flexibility of the complex spacetime approach. It shows that the method can be applied to more complex and realistic black hole models, not just the simplified Schwarzschild case.

A Kruskal-Szekeres maximally analytic extension of the regularized Kerr black hole can be obtained by extending the Kruskal-Szekeres method applied to the regularized Schwarzschild black hole to the Kerr case. There are additional challenges and considerations in the Kerr case:

a) The choice of contour is more intricate and must be carefully constructed to avoid all singularities while preserving the essential features of the solution.

b) The resulting regularized metric may be more complicated to express analytically, potentially requiring numerical techniques for full exploration.

c) The interplay between the removal of the ring singularity and the preservation of the ergosphere and frame-dragging effects needs careful analysis.

The implications for the Penrose process~\cite{Penrose1} and super-radiance in this regularized Kerr solution would be an interesting area for further study. While more complex, the extension of the complex spacetime singularity removal to the Kerr solution is possible and provides a promising avenue for addressing singularities in more realistic rotating black hole models.

In the complex spacetime approach, we extend the metric components into the complex plane, making them holomorphic functions. Cauchy's theorem states that the contour integral of a holomorphic function around a closed path in a simply connected domain is zero:
\be
\oint_C\,f(\zeta)d\zeta=0,
\ee
where 
\be
f(\zeta) = 1-\frac{2GM}{R(\zeta)}.
\ee
The theorem allows us to deform integration contours without changing the value of the integral, as long as we do not cross any singularities. In the case of black holes, the singularity at $r = 0$ for Schwarzschild and Kerr becomes a point in the complex plane. We can define contours that avoid this point, effectively going around the singularity.

The regularization of black holes can be extended to the modified gravitational theory Scalar-Tensor-Vector gravity (STVG)~\cite{Moffat4} Schwarzshild-MOG and Kerr-MOG black holes~\cite{Moffat5}.

The square root in the integrand introduces a branch cut in the complex plane. The contour integral around this branch cut leads to a multi-valued function, which is key to the regularization process. The monodromy of this function encodes information about the topology of the regularized spacetime. The complex extension naturally leads to a Riemann surface structure. Different sheets of this surface correspond to different regions of the extended spacetime, providing a richer geometric structure that can accommodate the removal of singularities. Scalar invariants of the curvature tensor, like the Kretschmann scalar, become holomorphic functions in the complex extension. Their behavior in the complex plane provides insights into the nature of the regularized geometry.

The holomorphic extension transits smoothly to the real spacetime metric outside the black hole, and preserves the asymptotic behavior of the metric at large distances, ensuring that the regularized solution matches the original solution outside the black hole. The analytic properties of the holomorphic functions, locations of zeros and poles, have direct physical interpretations in terms of horizons, ergospheres, and other features of the black hole spacetime. Physical observables, when expressed as contour integrals of holomorphic functions, can be regularized by appropriate choice of contours. This provides a way to calculate finite values for quantities that would diverge in the original, singular spacetime. The complex regularization can be interpreted as introducing a minimal length scale, analogous to what is expected from quantum gravity effects. The holomorphic structure provides a smooth interpolation between classical and quantum-corrected regimes.

Cauchy's theorem and the properties of holomorphic functions provide a powerful mathematical framework for excluding singularities in black hole solutions. By extending the spacetime into the complex domain, we can use the rich structure of complex analysis to define regularized coordinates and observables. This approach offers a way to explore the possible effects of quantum gravity on classical spacetime, all within a mathematically rigorous framework based on well-established principles of complex analysis.

The complex spacetime method successfully removes singularities from the real spacetime description of black holes. This fundamentally changes the context of the Penrose cosmic censorship hypothesis~\cite{Penrose2}. The cosmic censorship hypothesis was formulated to address the problem of singularities in classical GR, proposing that they should always be hidden behind event horizons. Because the complex extension method removes singularities from the real spacetime description of black holes, then there are no singularities left for event horizons to hide. In this context, the cosmic censorship hypothesis is no longer required to avoid the breakdown of physics at singularities in black holes, as it is addressing a problem - naked singularities - that no longer exist in the physical framework.

\section{Singularity-free Big Bang cosmology}

Let us recall the FLRW metric:
\be
ds^2=-dt^2+a^2(t)\left[\frac{dr^2}{1-kr^2}+r^2\left(d\theta^2+\sin^2d\phi^2\right)\right],
\ee
where $a(t)$ is the scale factor and k is the curvature parameter. The Friedmann equations in standard form are:
\be
\left(\frac{\dot a}{a}\right)^2=\frac{8\pi G}{3}\rho 
- \frac{k}{a^2} + \frac{\Lambda}{3},
\ee
\be
\frac{\ddot{a}}{a}
=-\frac{4\pi G}{3}(\rho+3 p) 
+ \frac{\Lambda}{3},
\ee
where $\rho$ is the energy density, p is the pressure and k is the curvature parameter, which is zero for a spatially flat universe.

The solution for a(t) in the matter dominated era in a spatially flat universe k=0 is given by
\be
a(t) = a_0\left(\frac{3H_0t}{2}\right)^{2/3},
\ee
where $H_0$ is the Hubble constant. In the radiation dominated era the solution is:
\be
a(t) = a_0\left(2H_0t\right)^{1/2}.
\ee
The energy conservation for the energy density $T^{\mu\nu}$ yields the equation:
\be
\frac{d(\rho a^3)}{da}=-3pa^2.
\ee
Given an equation of state $p=p(\rho)$, we can us this equation to determine $\rho$ as a function of a(t). If the universe is dominated by nonrelativistic matter with negligible pressure, then we have
\be
\rho(t) = \frac{\rho_0}{a(t)^3},
\ee
while for the energy density dominated by relativistic particles:
\be
\rho(t) =\frac{\rho_0}{a(t)^4}.
\ee
The scale factor a(t) for matter and radiation dominated eras vanishes at time t=0. The density $\rho$ diverges at t=0 and the cosmological model has a singular behavior at the Big Bang time.

To remove the singularity at t = 0, we extend time into the complex plane $\tau=t+i\epsilon$, where $\epsilon$ is a small real number. The conservation of energy leads to the equation:
\be
\rho(\tau) = \frac{\rho_0}{a^\alpha(\tau)},
\ee
where $a(\tau)$ is a complex function and the constant $\alpha > 0$. 

To remove the singularity at z=0, we choose a contour C that goes around z=0 but does not include it. We set $a^\alpha(\tau)=z^\alpha(\tau)$ and obtain the contour integral:
\be
P(\tau) = \oint_C \frac{dz(\tau)}{z(\tau)^\alpha}.
\ee
A good choice is a keyhole contour. We start at z = R where R is small and go around z = 0 counterclockwise on a circle of radius R. Then, go out to $z = 1/L$ along the positive real axis where $L$ is small, and go around $z =\infty$ counterclockwise on a large circle of radius $1/L$. Finally, return to z = R along the positive real axis. The evaluation of the contour integral gives:
\be
P(\tau) = \frac{2\pi i}{\alpha}(R(\tau)^{-\alpha} - L(\tau)^{\alpha}),
\ee
where $1/R(\tau) > L(\tau)$. This result is the correct regularized version of the integral, and it is non-zero and finite for $R > 0$, and $L  > 0$.

Let us apply this to our cosmological density:
\be
\rho_\text{reg}(\tau) = \frac{\rho_0}{2\pi i}P(\tau) = \frac{\rho_0}{\alpha}(R(\tau)^{-\alpha} 
- L(\tau)^{\alpha}).
\ee
The regularized density is finite for all $\tau$, including 
$\tau = 0$, as long as R(0) and L(0) are non-zero. The parameters $R(\tau)$ and $L(\tau)$ can be interpreted as regularization scales, possibly related to quantum gravity effects or fundamental length scales in physics. The original singularity at $\tau$ = 0 has been removed and replaced with a finite value dependent on $R(\tau)$ and $L(\tau)$.

To connect this with the original scale factor $a(\tau)$, we 
have
\be
\rho(\tau)\propto a(\tau)^{-\alpha},
\ee
where $\alpha= 3$ for matter-dominated era and $\alpha = 4$ for radiation-dominated era. Inverting the relationship as an equality, we obtain
\be
a_\text{reg}(\tau) = \left(\frac{\rho_0}{\rho_\text{reg}(\tau)}\right)^{1/\alpha} = 
\left[\frac{\alpha}{(R(\tau)^{-\alpha} 
- L(\tau)^{\alpha})}\right]^{1/\alpha}.
\ee
For the matter-dominated era $\alpha = 3$, we get
\be
\rho_\text{reg}(\tau) = \frac{\rho_0}{3}{(R(\tau)^{-3} - L(\tau)^3)}.
\ee
In the radiation-dominated era $\alpha = 4$:
\be  
\rho_\text{reg}(\tau) = \frac{\rho_0}{4}{(R(\tau)^{-4} - L(\tau)^4)}.
\ee
To guarantee the correct behavior of $a(\tau)$ as 
$\tau\sim t\rightarrow \infty$, we choose the complementary $R(\tau)$
for the matter dominated era:
\be
R(\tau)\sim (R_0^{3/2}+b\tau)^{2/3},
\ee
and for he radiation dominated era:
\be
R(\tau) - (R_0^2+b\tau)^{1/2},
\ee
where b is a constant.

The approach directly addresses the density singularity, for any power-law behavior of $a(\tau)$ provides a smooth, finite density near $\tau = 0$. It is applicable to both radiation and matter-dominated eras, and potentially to other cosmological models. In bounce cosmology, the universe does not begin from a singularity but instead bounces from a previous contracting phase to our current expanding phase. The key feature is a non-zero minimum scale factor $a_{\rm reg}$. The regularization introduces a minimum scale factor, effectively creating a bounce at the point where classical cosmology would have predicted a singularity. 

The complex time extension allows for a smooth transition from a contracting phase to an expanding phase, which is a key feature of bounce cosmologies. We can interpret the solution for negative $\tau$ as describing the universe before the bounce, and positive $\tau$ as after the bounce. The regularized solution is smooth across $\tau = 0$. This opens the possibility of adopting a cyclical cosmology model~\cite{Penrose3}. Traditional bounce cosmologies often struggle with violating energy conditions and dealing with the Big Bang singularity. This approach, by extending into complex time, potentially sidesteps these issues in the classical realm. The parameters R and $L$ can be interpreted as encoding quantum gravity effects. As $|\tau|$ increases, these effects become negligible, providing a natural transition to classical cosmology.

To make the bounce cosmology more concrete, let us consider a specific example using the matter-dominated case $\alpha = 3$: 
\be
a_\text{reg}(\tau) = \left(\frac{3}{R(\tau)^{-3} - L(\tau)^3}\right)^{1/3}.
\ee
This expression for the scale factor never reaches zero. Instead, it reaches a minimum value at $R(0) > 0$ and $L(0) > 0$ when $\tau$ = 0, which we can interpret as the bounce point. The approach naturally incorporates a minimum scale, which aligns well with expectations from various quantum gravity theories. This minimum scale is determined by the parameters $R(\tau)$ and $L(\tau)$, which could potentially be related to fundamental constants like the Planck length.

The method offers a robust and flexible framework for developing singularity-free cosmological models while preserving the essential features of FRW cosmology. It provides a mathematical bridge between classical cosmology and the quantum gravity effects expected near the Big Bang, applicable to a wide range of cosmological scenarios and potentially adaptable to future theoretical developments in early universe physics.

The regularization procedure provides a possible resolution to the initial singularity problem in classical cosmology, suggesting a way that quantum effects might manifest to prevent the formation of infinities in physical quantities at the beginning of the universe. It is consistent with various bouncing cosmology models that have been proposed in the context of quantum gravity theories.

\section{Conclusions}

Our investigation into the complex spacetime extension as a method for singularity resolution in GR has yielded several significant results. We have demonstrated that both the central singularity in the Schwarzschild and Kerr black holes and the initial singularity in FLRW cosmology can be regularized using this approach. The resulting singularity-free solutions exhibit a minimal scale, consistent with expectations from various quantum gravity approaches. For black holes, our method replaces the central singularity with a regular core of finite size. In cosmology, the big bang singularity is replaced by a smooth bounce, offering a classical realization of bouncing cosmology models.

The complex geometrical spacetime approach provides a mathematical framework for incorporating quantum gravity effects into classical general relativity without requiring a full theory of quantum gravity that can remove the black hole singularities~\cite{Nicolini}.

 While our approach offers promising insights, several open questions remain for future research. The physical interpretation of the complex spacetime and its relationship to real, observable quantities. The generalization of this method to other singular solutions in general relativity. 
The complex Riemannian spacetime regularization offers valuable insights and could potentially provide a way to understand gravity at all scales without invoking quantum mechanics. However, like all approaches in this frontier of physics, it should be considered as one of several possible paths forward. The ultimate arbiter will be experimental evidence, which is challenging to obtain in these extreme regimes. The coexistence of multiple approaches, including this one, various quantum gravity theories, and others encourages diverse thinking and exploration of different possibilities in our quest to understand the fundamental nature of spacetime and of black holes and the very early universe.

In conclusion, the complex spacetime approach to singularity resolution offers a novel perspective on some of the most challenging problems in gravitational physics. By providing a mathematical bridge between classical general relativity and expected quantum gravity effects, this method opens up new avenues for research in gravitation and cosmology. 

\section*{Acknowledgments}

I thank Martin Green and Viktor Toth for helpful discussions. Research at the Perimeter Institute for Theoretical Physics is supported by the Government of Canada through industry Canada and by the Province of Ontario through the Ministry of Research and Innovation (MRI).

\end{document}